\documentclass{caps}

\usepackage{peter}
\usepackage{epsf}

\usepackage{times}
\usepackage{mathptm}

\def\msun{$M_{\odot}$ }
\def\etal{{\em et al.} }
\def\bm#1{\mbox{\boldmath{$#1$}}}

\begin{document}

\pagenumbering{arabic}

%\deepchaptertrue    %If author is more than 2

\author{David L. Meier and Masanori Nakamura\\
Jet Propulsion Laboratory, California Institute of Technology, \\
Pasadena, CA 91109
}

\chapter{MHD Supernova Jets:  The Missing Link}

\begin{abstract}%
We review recent progress in the theory of jet production, with particular
emphasis on the possibility of 1) powerful jets being produced in the
first few seconds after collapse of a supernova core and 2) those jets
being responsible for the asymmetric explosion itself.  The presently
favored jet-production mechanism is an electrodynamic one, in which
charged plasma is accelerated by electric fields that are generated by a
rotating magnetic field anchored in the protopulsar.  Recent observations
of Galactic jet sources provide important clues to how {\em all} such
sources may be related, both in the physical mechanism that drives the
jet in the astrophysical mechanisms that create conditions conducive to
jet formation.  We propose a grand evolutionary scheme that attempts to
unify these sources on this basis, with MHD supernovae providing the
missing link.  We also discuss several important issues that must be
resolved before this (or another scheme) can be adopted.
\end{abstract}

\section{Introduction: A Cosmic Zoo of Galactic Jet Sources}
\label{sec:intro}
The last few decades have seen the discovery of a large number of
different types of Galactic sources that produce jets.  The purpose
of this talk is to show that all of these jets sources are related,
in both a physical sense and an astrophysical sense.  Furthermore,
Craig Wheeler's idea that most core collapse supernovae (SNe) are driven by
MHD jets from a protopulsar provides the missing link in an attractive
unified scheme of all stellar jet sources.  Below is a list of Galactic
jet sources that have been identified so far:

\begin{enumerate}
\item Jets from Stars Being Born:  Protoplanetary systems.  These are
jetted and bi-polar outflows from young stars or star-forming regions.
They are associated with protostars that have protoplanetary disks and
are ejected at approximately the escape velocity of the central star
($v_{jet} \sim v_{esc} \sim 200 \, {\rm km \, s^{-1}}$).

\item Jets from Dying Stars:
\begin{itemize}
\item Planetary Nebula Systems.  Stars with initial mass less than a
few \msun end their lives as a planetary nebula, leaving a white dwarf
remnant.  Many PN have bipolar shapes; some even have highly-collimated
outflows of up to $v_{jet} \sim 1000 \, {\rm km \, s^{-1}}$, the escape
speed from the surface of a rather distended white dwarf.

\item Core-collapse Supernovae:  Stars with masses between roughly 10
and 30 \msun are destroyed in an explosion that is triggered by the
collapse of their iron cores.  They leave a neutron star/pulsar remnant.
There is growing evidence (see below) that core-collapse SNe
also produce jets with a power comparable to the explosion itself.
Expected speeds are $v_{jet} \sim 0.25 - 0.5 c$ --- the escape speed
from the new protoneutron star.

\item Isolated young pulsars (Crab, Vela, etc.):  Jets now have been
detected by Chandra in these objects and have speeds $v_{jet} \sim 0.5 c$.
These may be the remaining ghosts of core-collapse jets that occurred 
thousands of years ago.

\item Gamma-ray bursts (GRBs):  Believed to be black holes in formation,
these produce jets with $\Gamma_{jet} \sim 100 - 300$ that point toward
us.  Long-duration GRBs are closely associated with powerful SNe
and may represent the death of particularly massive ($>$ 30 \msun) stars.
\end{itemize}

\item Jets from Re-kindled Dead Stars in Binary Systems
\begin{itemize}
\item Symbiotic stars ({\it e.g.}, R Aquarii):  These tend to be accreting
white dwarfs where the companion star has recently left the main sequence
and expanded into a red giant, transferring mass to the compact object
via a strong wind or Roche lobe overflow.  Jets are produced by the white
dwarf and have speeds up to $v_{jet} \sim 6000 \, {\rm km \, s^{-1}}$,
the escape speed from a white dwarf.

\item Neutron star X-ray Binaries:  These occur in both low-mass and
high-mass systems.  Jets tend to appear at lower accretion rates.
Typical speeds are $\sim 0.25 c$, but can reach speeds approaching $c$.

\item SS433-type objects:  Observed properties suggest a super-Eddington
accreting, magnetized neutron star in a high-mass binary system.
Jets have speed $v_{jet} \sim 0.25 c$.

\item Classical microquasars (GRS 1915+105, GRO J1655-40, GX 339-4, etc.):
These produce jets with $v_{jet} > 0.6 - 0.95 c$ ($\Gamma_{jet} \equiv
[ 1 - v_{jet}^{2} / c^{2} ]^{-1/2} > 1.25 - 3$) and up.  Virtually all
are black hole candidates in low-mass X-ray binary systems.
\end{itemize}
 
\end{enumerate}

The inclusion of core-collapse SNe above is the key to the
unified model presented below.  In the mid 1990s it was discovered
that such SNe emit polarized light in the optical band,
caused by electron scattering by an asymmetrically-expanding explosion
\cite{wang01,wang03,leo01}.  The variation of polarization properties with
time and with different SN types gives important clues to its nature.
In a given SN, the degree of polarization $\Pi$ increases with time
but the polarization direction remains constant in time and wavelength,
indicating that the asymmetry is global and maintains a fixed direction.
For Type IIa SN (ones with a large hydrogen-rich envelope), the $\Pi
\sim 1\%$, indicating only a 2:1 or less asymmetry.  For Type IIb SN
(ones that have lost their hydrogen envelope prior to the explosion,
leaving only the helium envelope), $\Pi$ is higher ($\sim$2\%), indicating
a 2.5:1 axial ratio.  For SN Type Ib/Ic (ones that have lost most or
all of their envelope, leaving a compact blue Wolf-Rayet star that then
explodes), $\Pi$ is quite high ($4-7$\%), indicating a 3:1 axial ratio
or better.  Clearly, the deeper one sees into the explosion, the more
elongated the exploding object appears.  These observers have concluded
that core-collapse SN have a global prolate shape that appears to be
associated with the central engine producing the explosion.  A jet,
with energy comparable to that of the explosion itself, significantly
alters the shape of the envelope, creating the elongated, polarized
central source.  In the paper below we will show that all of the above
sources may be intimately related, both in the physical origins of the
jet itself and in their astrophysical origins as well.

\section{Basic Principles of Magnetohydrodynamic Jet Production}
\label{sec:basic_principles}
The basic principles of MHD jet production have been described elsewhere
\cite{meier01}. The reader is referred to that paper for a more detailed
description, and more comprehensive figures, than in the short review
given below.

\subsection{Launching of the Jet Outflow:  The Jet Engine Itself}
\label{sec:jet_launching}

\subsubsection{Jet Production in Accreting Systems and Pulsars} 
\label{sec:generic_mhd_jets} 
Several mechanisms for producing bipolar outflows have been
suggested (explosions in the center of a rotationally-flattened cloud,
radiation-pressure-driven outflows from a disk, etc.), but none of these
is able to produce outflows approaching the highly-relativistic speeds
observed in the fastest jet sources.  The currently-favored mechanism
is an electro-magneto-hydrodynamic (EMHD) one, somewhat similar to
terrestrial accelerators of particle beams.  Indeed, electromagnetic
acceleration of relativistic pulsar winds has been a leading model for
these objects since the 1960s.  EMHD jet production was first suggested in
1976 \cite{bland76,love76} and has been applied model to rotating black
holes \cite{bz77} (BZ) and to magnetized accretion disks \cite{bp82}
(BP).  This mechanism has now been simulated and is sometimes called the
``sweeping pinch'' mechanism \cite{su85,kud99,nak01}.

The most important ingredient in the EMHD mechanism is a magnetic field
that is anchored in a rotating object and extends to large distances
where the rotational speed of the field is considerably slower.  Plasma
trapped in the magnetic field lines is subject to the Lorentz (${\bm J}
\times {\bm B}$) force, which, under conditions of high conductivity (the
MHD assumption), splits into two vector components: a magnetic pressure
gradient ($- {\bm \nabla} B^{2} / 8 \pi$) and a magnetic tension (${\bm B}
\cdot {\bm \nabla} {\bm B} / 4 \pi$).  Differential rotation between the
inner and outer regions winds up the field, creating a strong toroidal
component ($B_{\phi}$ in cylindrical [$R$, $Z$, $\phi$] coordinates).
The magnetic pressure gradient up the rotation axis ($-d B_{\phi}^{2}
/ d Z$) accelerates plasma up and out of the system while the magnetic
tension or ``hoop'' stress ($-B_{\phi}^{2} / R$) pinches and collimates
the outflow into a jet along the rotation axis.

This basic configuration of differential rotation and twisted magnetic
field accelerating a collimated wind can be achieved in all objects
identified in Sect. \ref{sec:intro}. For protostars, white dwarfs, X-ray
binaries, classical microquasars and GRBs, the field will be anchored
in the accreting plasma, which may lie in a rotating disk (BP) and/or
may be trapped in the rotating spacetime of the spinning central black
hole itself (BZ).  In the case of SS433-type objects, isolated pulsars,
and core-collapse SN the rotating field is anchored in the pulsar (or
protopulsar).  In SS433 and core-collapse SN the source of the accelerated
plasma is, once again, accretion, but in isolated pulsars it is believed
to be particles created in spark gaps by the high ($10^{12}$ G) field.

\subsubsection{Jets from Kerr Black Holes:\\
Direct \& Indirect Magnetic Coupling}
\label{sec:blackhole_mhd_jets}
The jet-production mechanism envisioned by BZ generally involved direct
magnetic coupling of the accelerated plasma to the rotating horizon.
That is, magnetic field lines thread the horizon, and angular momentum
is transferred along those field lines to the external plasma via
magnetic tension.  However, another, indirect, coupling is possible.
This mechanism, suggested by \cite{pc90} (PC) and recently simulated by
us \cite{koide02}, has the same effect as the BZ mechanism (extraction
of angular momentum from the rotating black hole by the magnetic field),
but the field lines do not have to thread the horizon itself.  Instead,
they are anchored in the accreting plasma.  When this plasma sinks into
the ergosphere near the black hole ($R < 2~G M / c^{2}$), frame dragging
causes the plasma to rotate with respect to the exterior, twisting up
the field lines in a manner similar to the situation when the field
is anchored in a disk or pulsar.  (This occurs even if the accreting
plasma has no angular momentum with respect to the rotating spacetime.)
The twisted field lines then have two effects:

\begin{enumerate}
\item Electromagnetic power is ejected along the rotation axis in the form
of a torsional Alfven wave.  Eventually the output Poynting flux power
should be dissipated in the production and acceleration of particles
and a fast jet.

\item The back-reaction of the magnetic field accelerates the ergospheric
plasma (in which it is anchored) to relativistic speeds {\em against}
the rotation of the black hole.  The counter-rotating ergospheric plasma
now formally has negative angular momentum and negative energy (negative
mass); that is, it has given up more than its rest mass in energy to the
external environment.  It is on orbits that must intersect the black
hole horizon, and, when it does, the mass of the black hole decreases
by a value equal to that negative energy.  \end{enumerate}

This process is the magnetic equivalent of the Penrose process,
but instead of extracting black hole rotational energy by particle
scattering, the energy is extracted by scattering of an Alfven wave
off the ergospheric plasma particles.  Determining whether the BZ or PC
process occurs in certain systems is an important question for future
study.

\subsection{Acceleration and Collimation (A \& C)}

\subsubsection{Slow A \& C is Probably the Norm}
\label{sec:slow_acc_normal}
There are both theoretical and observational reasons for believing that
slow acceleration and collimation is probably the norm for jet outflows
in these sources.  Non-relativistic \cite{kras99} and relativistic
\cite{vk01} models of MHD wind outflows attain solutions where the wind
opening angle is wide near the accretion disk and then narrows slowly
over several orders of magnitude in distance from the disk.  Because the
dynamical time scale is of order 0.1 ms or less in these objects, a
steady state is set up fairly quickly in jet ejection events that last
even only a few seconds.  In a steady state, the wind accelerates as it
expands vertically away from the rotator.  A jet is not fully formed
until its speed exceeds the local wave propagation speed, {\it i.e.},
the total Alfven speed $V_{A} = [(B_{R}^{2}+B_{Z}^{2}+B_{\phi}^{2}) /
(4 \pi \rho)]^{1/2}$, where $\rho$ is the mass density in the outflowing
material.  The place where this occurs, often called the Alfven point or
Alfven surface, generally is well above the rotating object producing
the accelerating torsional Alfven wave.  Analytic \cite{bp82,li92}
and numerical \cite{kras99} studies of this steady state show that the
outflow is rather broad at the base, and it slowly focuses as it is
accelerated. At a height $Z_{A} > 10 R_{0}$ above the disk, the total
Alfven speed is exceeded, the flow is focused into a narrow cylindrical or
conical flow, and little more acceleration and collimation takes place.
The terminal jet speed $v_{jet}$ is of order $V_{A}(Z_{A})$, and this
speed is usually of order the escape speed from the central rotator
$V_{esc}(R_{0})$.

Furthermore, there now is observational evidence that, in at least
some {\em extragalactic} systems, the steady-state picture of slow
acceleration and collimation is correct.  Very high resolution VLBA
radio images of the M87 jet \cite{junor99} show a broad $60\deg$ opening
angle at the base that narrows to only a few degrees after a few hundred
Schwarzschild radii.  In addition, it has been argued \cite{sm01} that
most quasar jets must be broad at the base:  they lack soft, Comptonized
and relativistically-boosted X-ray emission that would be expected from
a narrow, relativistic jet flow near the black hole.

\subsubsection{Stability of Highly-Magnetized Flows During A \& C}
\label{sec:pfd_stability}
Because the acceleration and collimation is expected to be slow, the
terminal velocity and final state of the outflow will be reached only
far from the central engine.  The character of the outflow, therefore,
will depend crucially on how it interacts with the external medium that
surrounds it in the acceleration region.  It is important, therefore,
to consider the effects of the ambient ``weather'' surrounding the jet.

Highly-magnetized flows are characterized by a high ratio of Alfven to
sound speed $c_{s}$
\begin{eqnarray}
\frac{\dot{\varepsilon}_{fields}}{\dot{\varepsilon}_{particles}} & = & 
\frac{v_{jet} \, B^2 / 4 \pi}{v_{jet} \, \rho c_{s}^2} \; \approx \; 
\left(\frac{V_{A}}{c_{s}} \right)^2 \; >> \; 1
\end{eqnarray}
This means that, early in the flow the velocity becomes supersonic,
but continues to accelerate toward the Alfven speed.  In this region,
where $c_{s} < v_{jet} < V_{A}$, the ratio of Poynting flux to
kinetic energy flux is
\begin{eqnarray}
\frac{\dot{\varepsilon}_{Poynting}}{\dot{\varepsilon}_{kinetic}} & = & 
\frac{ \frac{c}{8 \pi} | \bm{E} \times \bm{B} | }
{ v_{jet} \frac{1}{2} \, \rho {v_{jet}}^2} \; = \; 
\frac{B^2}{4 \pi \rho {v_{jet}}^2} \; = \; 
\left( \frac{V_{A}}{v_{jet}} \right)^2 \; > \; 1
\end{eqnarray}
That is, the flow is ``Poynting-flux-dominated'' as long as the flow
remains supersonic but sub-Alfvenic.

While semi-analytic models of such ``Poynting-flux-dominated'' jets have
been built \cite{li92,love02}, no numerical simulations of highly {\em
relativistic} jets have been performed yet.  The best numerical results
so far are from {\em non}-relativistic simulations \cite{nak01,nak03},
which compute the behavior of a jet in a decreasing density (increasing
$V_{A}$) atmosphere.  They show that the electromagnetic power is carried
by a fairly coherent torsional Alfven wave that encompasses the jet in
a twisting spiral pattern.  This wave can transport energy and momentum
along the flow, causing further acceleration far from the central engine.

\begin{figure}
\begin{center}
\leavevmode\epsfxsize=12cm \epsfbox{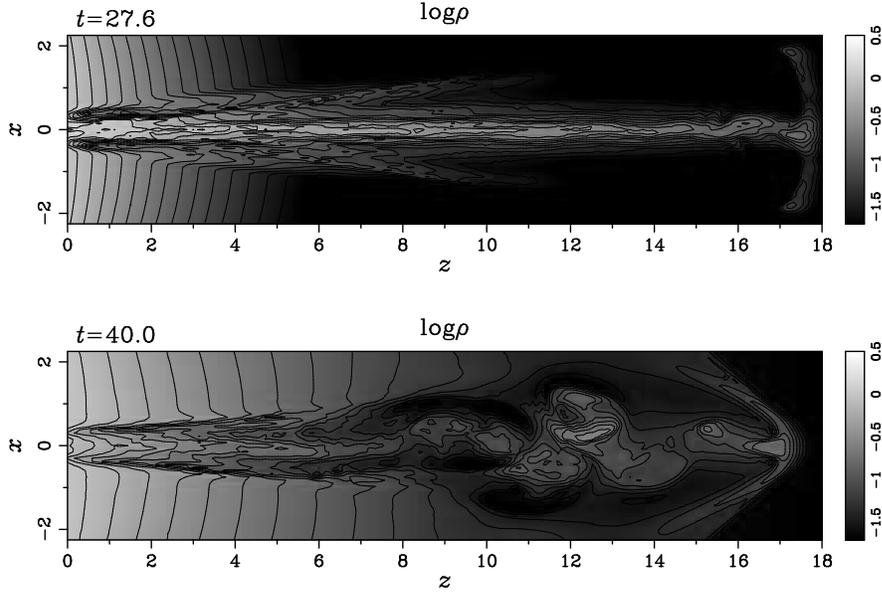}
\end{center}
%Figure caption
\caption{
Three-dimensional simulations of jets that are stable (top) and unstable (bottom) 
to the helical kink instability.  Both have $\beta_{plasma} < 1$ throughout 
the domain, but the stable model has $d \ln \beta_{plasma} / d \ln r < \sim 0$ 
($-d \ln \rho / d \ln r \sim 3$, {\it vs.} $2$ for the unstable model). 
(After \cite{nak03}.)
}
\label{fig:jet_simulations}.  
\end{figure}

Simulations of both stable and unstable jets are shown in Figure
\ref{fig:jet_simulations}.  We find that the stability of Poynting
flux-dominated flows is critically dependent on how severe the mass
entrainment in the jet is --- specifically on the {\em gradient}
of the plasma parameter $\beta_{plasma} \equiv p_{gas}/(B^2/8\pi)$.
(In the following, remember that $\beta_{plasma}$ is always less than
unity for PFD jets, if the plasma is reasonably cold $P \leq \rho c^2$.)
If $\beta_{plasma}$ decreases or remains small as the jet propagates
outward (mass loading becomes even less or stays the same), then we
find that the PFD jet remains stable.  However, if $\beta_{plasma}$ {\em
increases} (entrains significantly more thermal material), then we find
that the jet is likely to be unstable to the helical kink instability,
{\em even if the jet still remains magnetically dominated throughout
the simulation}.  Apparently even a small amount of pressure in the flow
builds up over large distances, triggering a helical kink and, therefore,
turbulence in the jet.

Collimated MHD outflows that are strongly dominated by magnetic stresses,
and which remain that way or increase in strength, therefore propagate
as straight and stable jets. Those in which the relative strength
of the field decreases along the jet become unstable to the helical
kink instability.  In the latter case the outflow decelerates and
deposits its momentum in a broad cone rather than punching through the
ambient medium in a narrow jet.  $\beta_{plasma} << 1$ is, therefore,
a necessary, but not sufficient condition for jet stability.  We also
must have $d \ln \beta_{plasma} / d \ln r < \sim 0$.  This may have
important implications for MHD outflows from newly-formed pulsars in
the centers of core-collapse SNe.

\section{MHD-Jet-Powered Supernovae} 

\subsection{Basic MHD Supernova Model}

Several authors have suggested in the past that MHD phenomena may power
supernovae \cite{lw70,bk71}. The recent discovery that SN ejecta are
elongated by an asymmetric jet-like flow has stimulated renewed interest
in these models and, in particular, in the possibility that an MHD jet
produced by the protopulsar may be the source of the explosion energy
in all core-collapse SN.  We have proposed \cite{w02} an explosion
mechanism that is consistent with the above properties of MHD jets. The
jet is produced in the iron mantle, just outside the protoneutron star.
An object with a $10^{15}$ G field and a rotation period of $\sim 1$
ms can produce a jet power of $\sim 3 \times 10^{51} {\rm erg~s^{-1}}$
and a total energy of $\sim 2 \times 10^{52}$ erg --- more than enough to
eject the outer envelope and account for the observed explosion energy.
The ejecta from the vicinity of the core is nickel-iron rich, with a
mass of $\sim 10^{32} {\rm g}$, an initial velocity of $\sim 0.5 c$,
and a total momentum of $\sim 1.5 \times 10^{42} {\rm g \, cm \, s^{-1}}$, 
decelerating somewhat as it passes through the stellar envelope.

The protoneutron star spins down to more respectable rotation periods ($>
10$ ms) in about $\sim 10 \, B_{15}^{-2}$ seconds.  (The model is similar
to that of Ostriker \& Gunn \cite{og71}, with their $10^{12}$ G pulsar
fields replaced by $10^{15}$ G protopulsar fields.)  The jet outflow
is composed of iron-rich material and is initially broad at the base.
Furthermore, as even these field strengths do not satisfy the necessary
condition for jet stability, the jet likely will be subject to helical
kink instabilities, broadening further into a wide bipolar outflow
at large distances ($> 10^{7-8}$ cm) from the core.  It therefore can
couple well to the outer envelope and eject it, imparting an elongated
shape to the supernova explosion.

\subsection{The Pulsar Rocket}

The Crab pulsar currently has a proper motion of $\sim 120 \, {\rm
km \, s^{-1}}$ along the pointing direction of one of its twin jets.
This motion has an energy of $\sim 2 \times 10^{47} {\rm erg}$ and
momentum of $\sim 3 \times 10^{40} {\rm g \, cm \, s^{-1}}$.  However,
while the current Crab pulsar jets have sufficient energy to have powered
this motion ($\sim 10^{49} {\rm erg}$, if they had operated continuously
over the past 950 yr), the present Crab jet does not produce enough momentum to
accelerate the pulsar to this speed, {\em even if the jet had been continuously
and completely one-sided} (100\% asymmetric, or $10^{39} {\rm g \, cm \, s^{-1}}$).
On the other hand, the MHD jet inferred from the MHD supernova model
above would have been quite sufficient to supply both the energy
($\sim 10^{52} {\rm erg}$) and momentum with only a 2\% jet asymmetry.
The fact that the Crab jet and proper motion are aligned, indicates that
the current observed Crab pulsar jet (with a speed of $\sim 0.5 c$)
may be a vestige of the original $\sim 0.5 c$ jet that exploded the
supernova and accelerated the pulsar.

\subsection{A Gamma-Ray Burst Trigger}

This model also includes a gamma-ray burst trigger.  In very rare
instances, the field can be dynamically strong in the iron mantle ($B >
10^{16}$ G), leading to satisfaction of the necessary (but not sufficient)
jet stability condition $\beta_{plasma} << 1$.  If the density gradient
then is also steep ($d \ln \beta_{plasma} / d \ln r < \sim 0$) then
the jet becomes narrow and very fast, coupling poorly to the mantle,
punching through the outer envelope \cite{kh01}, escaping the star,
and producing a heavy iron ``lobe'' outside it, traveling at a speed of
0.05-0.3 c.  Because of the poor coupling to the mantle, the explosion fails, and much
of the mantle falls back onto the protoneutron star, putting the system
into a state very similar to that at the beginning the ``failed SN'' GRB
model of \cite{mw99}.  When the mantle fallback accretes enough material
onto the protoneutron star (after several minutes to hours), the neutron
star is crushed to a black hole, and a new very fast ($\Gamma_{jet}
>> 1$) jet is produced via the BZ or PC mechanisms discussed above.
The relativistic jet catches up with the slow, iron-rich lobe at a
distance of $d \sim v_{jet} \tau_{fallback} \sim 10^{12-13}$ cm, and
the interaction of jet and lobe produces gamma-rays, optical afterglow,
and an iron-rich spectrum.

\subsection{Unresolved Issues}

The $10^{14-15}$ G magnetic field strengths needed in SN cores are the
real key to the success of the MHD SN model.  Magnetars are believed
to have surface field strengths of this order, but pulsars typically
have fields of order $10^{12-13}$ G.  Fields of this strength would
have produced a slow supernova explosion lasting several months.  The
prediction, then, is that fast core-collapse SNe produce magnetars
and slow SNe produce pulsars, but there is no observational
evidence for (or against) this prediction, neither direct nor statistical.
On the other hand, if stronger fields {\em did} exist in pulsars at the
time they were formed, they must have been dissipated either during the
SN process or shortly thereafter, but it is not known how that dissipation
may have taken place. Secondly, there also may be competing jet mechanisms
(neutrino radiation pressure, etc.) which we have not discussed here.
Thirdly, while we have suggested a possible SN failure mechanism, much
more detailed theoretical work will be needed before we will be able to
perform the simulations necessary to test this and other such mechanisms.
Finally, there is a problem that needs to be addressed by {\em all} SN models.
The iron mantle in the progenitor star is very neutron rich.  If much
of it is ejected (and the $1.4 M_{\odot}$ protoneutron star is left),
then the predicted amount of {\em r}-process material may be much larger
than that observed.  It is a general problem for all core-collapse SN
models to produce a neutron star remnant while still not over-producing
the {\em r}-process elements.

\section{A Grand Unified Scheme for All Galactic Jet Sources}

SN and long-duration GRBs, therefore, potentially can be unified
astrophysically as being different possible outcomes in the final stages
of the death of a massive star.  Also, if the above model is applicable,
they both can be unified {\em physically} as being powered by MHD jets.
This strongly suggests, therefore, a possible {\em grand scheme} that
unifies all Galactic jet sources discussed in Sect. \ref{sec:intro},
both physically and astrophysically, in a similar manner.

\begin{figure}
\begin{center}
\leavevmode\epsfxsize=10cm \epsfbox{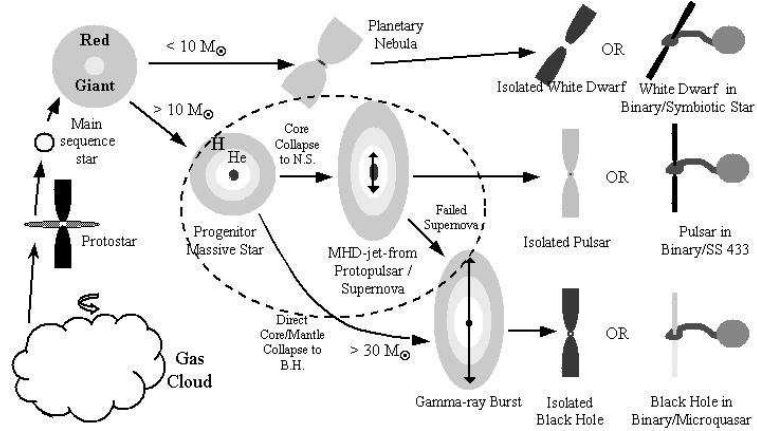}
\end{center}
%Figure caption
\caption{
The proposed grand unified model for Galactic jet sources.  (After
\cite{meier03}.)
}
\label{fig:unified_model}
\end{figure}

The properties of all Galactic jet sources are similar in several ways.
They all appear to produce jets when there is accretion, shrinkage, or
collapse of plasma in a gravitational field.  Because of that shrinkage
or collapse, they occur in systems that are probably in a rapid rotation
state:  conservation of angular momentum implies that even a modest
amount of rotation before the collapse would be amplified greatly.
They also are associated with systems that have strong magnetic fields.
Some directly reveal these magnetic fields in their radio synchrotron
emission.  Others are produced by stars that are believed to have strong
fields for other reasons (protostars, accretion disks, newly-formed
protoneutron stars).  The grand unified scheme, therefore, asserts
that all Galactic jet sources represent objects in which an excess of
angular momentum has built up because of accretion or collapse. And the
production of the jet itself represents the {\em expulsion} of that excess
angular momentum by electro-magneto-hydrodynamic processes.  The unifying
evolutionary sequence for all Galactic jet sources is shown in Figure
\ref{fig:unified_model}.

\subsection{Protostellar and White Dwarf Galactic Jets}

The unified sequence begins with the formation of a protostar in a
collapsing interstellar cloud. A jet is formed during the accretion
phase, and is responsible for spinning down the star to the relatively
low rotation rates seen on the main sequence.  In low-mass stars jet
production does not resume until shrinkage of the central stellar
proto-white-dwarf core produces a bipolar planetary nebula outflow.
Symbiotic star-type jets are expected in binary systems, but isolated
white dwarfs may have neither the accretion fuel nor the rapid spin to
do so.

\subsection{Neutron Star and Black Hole Jets}

The MHD supernova model discussed earlier provides the missing link to
GRB, X-ray binary, and microquasar jets.  In the grand unified scheme,
most massive star cores collapse to a neutron star, ejecting a broad
MHD jet in the process that drives the SN explosion and produces the
observed asymmetry.  After the envelope dissipates, if the pulsar is
an isolated object, residual rotation of the magnetized remnant still
drives an MHD outflow and a moderately relativistic jet like that seen
in the Crab and Vela pulsars.  On the other hand, if the pulsar resides
in a binary system, it may accrete material from its companion star in
a super-Eddington phase ($\dot{M}_{acc} >> 10^{18} {\rm g~s^{-1}}$)
and appear like SS433 for a brief time.  Cessation of the accretion,
angular momentum evolution of the pulsar, or possible collapse to a
black hole all could serve to alter SS433's present state.

In rare circumstances, the MHD SN jet will fail to eject the envelope, or
perhaps the progenitor core will collapse directly to form a black hole.
In either case a GRB event should be generated in a manner similar to
that in the failed SN model.  The GRB jet event will not spin down the
black hole completely (although a significant amount of rotational energy
may be extracted from the black hole by another means --- gravitational
waves).  After the envelope dissipates, if the black hole is an isolated
object, it may emit little radiation and be difficult to detect.  On the
other hand, if the newly-formed hole is in a binary system, it also can
accrete plasma and field from its companion, thereby producing a strong
jet and classical microquasar.  Again, changes in the accretion rate and
angular momentum evolution of the black hole will alter the microquasar's
observational state. 

The key element of this unified model is that the evolutionary outcome
of a star is ultimately determined by the magnitude and direction of the
angular momentum and magnetic field of that star core's.  We therefore
should consider the jets seen in young pulsars and SS433-type objects to
be vestiges of the mechanism that exploded the massive stars from which
they came and, in a similar manner, consider the jets in microquasars to
be the remnants of the gamma-ray burst that triggered the black hole's
formation eons ago.  These relatively modest jets are the echoes of
violent events of the distant past.

The research described in this paper was carried out at the Jet Propulsion
Laboratory, California Institute of Technology, under contract to the
National Aeronautics and Space Administration.  M.N. was supported, in 
part, by a National Research Council Resident Research fellowship from 
NASA.

\end{document}